\documentclass[journal=jpccck,manuscript=article]{achemso}

\usepackage[version=3]{mhchem} 
\usepackage[T1]{fontenc}       
\usepackage[usenames]{color}
\usepackage{graphicx}
\usepackage{float}

\author{Ana M. Valencia}
\affiliation[Humboldt-Universit\"{a}t zu Berlin] {Institut f\"{u}r Physik und IRIS Adlershof, Humboldt-Universit\"{a}t zu Berlin, 12489 Berlin, Germany}
\email{ana.valencia@physik.hu-berlin.de}
\author{Caterina Cocchi}
\affiliation[Humboldt-Universit\"{a}t zu Berlin] {Institut f\"{u}r Physik und IRIS Adlershof, Humboldt-Universit\"{a}t zu Berlin, 12489 Berlin, Germany}
\email{caterina.cocchi@physik.hu-berlin.de}

\title{Electronic and Optical Properties of Oligothiophene-F4TCNQ Charge-Transfer Complexes: The Role of Donor Conjugation Length}

\begin{document}
\begin{abstract}
We investigate from first-principles many-body theory the role of the donor conjugation length in doped organic semiconductors forming charge-transfer complexes (CTCs) exhibiting partial charge transfer. 
We consider oligothiophenes (nT) with an even number of rings ranging from four to ten, doped by the strong acceptor 2,3,5,6-tetrafluoro-7,7,8,8-tetracyano-quinodimethane (F4TCNQ). 
The decrease of the electronic gaps upon increasing nT size is driven by the reduction of the ionization energy with the electron affinity remaining almost constant.
The optical gaps exhibit a different trend, being at approximately the same energy regardless of the donor length.
The first excitation retains the same oscillator strength and Frenkel-like character in all systems.
While in 4T-F4TCNQ also higher-energy excitations preserve this nature, in CTCs with longer nT oligomers charge-transfer excitations and Frenkel excitons localized on the donor appear above the absorption onset.
Our results offer important insight into the structure-property relations of CTCs, thus contributing to a deeper understanding of doped organic semiconductors.
\end{abstract}
\newpage
\section{Introduction}
Doping in organic semiconductors is an ubiquitous phenomenon appearing when donor and acceptor molecules are combined together, and crucially determines the electronic, optical, and transport properties of the resulting materials~\cite{aziz+07am,yim+08am,ping+10jpcl,gao+13jmcc,ping-nehe13prb,duon+13oe,salz+13prl,mend+13acie, yang+14ami, mend+15ncom,salz+16acr,fuze+16jpcl,jaco-moul17am,duva+18jpcc,beye+19cm}.
Two main mechanisms have been identified to be responsible for doping in organic semiconductors: Integer charge transfer (ICT) manifests itself when an electron is transferred from the donor to the acceptor, leading to the formation of ion pairs. 
Partial charge transfer occurs upon electronic hybridization of the frontier orbitals of the donor and the acceptor, giving rise to a charge-transfer complex (CTC). 
While a general consensus has been established around these two processes, recent studies showing that ICT and CTC can coexist in the same molecular material~\cite{jaco+18mh,neel+18jpcl} have reignited the debate about doping in organic semiconductors.
One of the open questions in this respect concerns the role of the donor conjugation length in determining the electronic structure and the optical properties of these systems.
A recent study by Mendez \textit{et al.}~\cite{mend+15ncom} shows that ICT appears when the strong electron acceptor 2,3,5,6-tetrafluoro-7,7,8,8-tetracyano-quinodimethane (F4TCNQ) is blended with the poly-3-hexylthiophene (P3HT) polymer, (see also Refs.~\citenum{yim+08am,ping-nehe13prb,duon+13oe,karp+17mm,scho+17afm}), while CTCs are formed when the same acceptor is coupled to the quarterthiophene oligomer.
These findings suggest a fundamental relation between the doping mechanism and the length of the donor chain.
On the way to tackle this problem in its general complexity, it is relevant to first understand how the conjugation length of the donor oligomers impacts the electronic structure and the optical excitations of complexes exhibiting only partial charge transfer.
Rationalizing this behavior in the presence of one specific doping mechanism is essential to set the stage for further theoretical and experimental studies.

We perform a first-principles study based on density-functional theory (DFT) and many-body perturbation theory (MBPT), which offer state-of-the-art accuracy and insight into the electronic and optical properties of the investigated systems~\cite{blas-atta11apl,baum+12jctc,fabe+14ptrsa,nied+15afm}.
We focus on CTCs formed by thiophene oligomers (nT) $p$-doped by F4TCNQ.
We consider thiophene chains with an even number of rings ranging from 4T, which is the shortest oligomer typically investigated in experiments~\cite{aziz+07am,ping+10jpcl,mend+15ncom}, up to 10T.
All systems are modeled as molecular dimers \textit{in vacuo}~\cite{zhu+11cm,gao+13jmcc,mend+13acie,mend+15ncom}, with the donor and the acceptor molecules in the $\pi$-$\pi$ stacking arrangement~\cite{zhu+11cm,gao+13jmcc}.
In this way, we are able to systematically rationalize the role of the donor length on the electronic and optical properties of CTCs which differ from each other only by the size of the nT oligomers.
Starting from the level alignment of the individual components, we relate the band-gap reduction upon increasing donor length to the variation of the ionization energy of the system.
In the analysis of the optical properties we examine how the absorption onset and the spectral density are affected by the size of the nT oligomer.
We finally discuss the character of the lowest-energy excitations in terms of electron and hole density and demonstrate that their behavior is critically sensitive to the donor conjugation length.

\section{Theoretical Background and Computational Details}
Equilibrium geometries of the individual donor and acceptor molecules as well as of the CTCs are computed from DFT with the all-electron package FHI-aims~\cite{blum+09cpc} adopting tight integration grids, TIER2 basis sets~\cite{havu+09jcp}, and the Perdew-Burke-Ernzerhof~\cite{perd+96prl} generalized-gradient approximation for the exchange-correlation functional.
Van der Waals interactions, that play an essential role in $\pi$-$\pi$ stacked systems, are accounted for by means of the Tkatchenko-Scheffler scheme~\cite{tkat-sche09prl}. 
Atomic positions are relaxed until the Hellmann-Feynman forces are smaller than 10$^{-3}$ eV/\AA{}.
The acceptor molecule adsorbed in the middle of the donor chain gives rise to the most stable configuration, in agreement with previous first-principles results~\cite{zhu+11cm}.
Further details and discussion about the geometry of the CTCs are reported in the Supporting Information. 

The electronic structure and the optical properties of the considered systems are computed with the MOLGW code~\cite{brun+16cpc}.
Gaussian-type cc-pVDZ basis sets~\cite{brun12jcp} are used including the frozen-core approximation. 
The resolution-of-identity approximation is also employed~\cite{weig+02}. 
The $GW$ approximation in the perturbative approach $G_0W_0$ is adopted to calculate the quasi-particle (QP) correction to the single-particle levels starting from the solution of the Kohn-Sham equations.
The range-separated hybrid functional CAM-B3LYP~\cite{yana+04cpl} is chosen to obtain a reliable starting point for the $G_0W_0$ calculations~\cite{fabe+13jcp,brun+13jctc}.
Optical absorption spectra and (bound) electron-hole pairs are computed from the solution of the Bethe-Salpeter equation (BSE), the equation of motion for the two-particle polarizability, $L = L_0 + L_0 \, K \, L$~\cite{brun+16cpc}.
The so-called BSE kernel $K$ includes the repulsive exchange term given by the bare Coulomb potential $\bar{v}$ and the statically screened electron-hole Coulomb attraction $W$. 
For further details about this formalism, see, \textit{e.g.}, Refs.~\citenum{fabe+13jcp,cocc-drax15prb,brun+16cpc}.
We solve the BSE in the Tamm-Dancoff approximation (TDA), which does not impact the nature of the excitations analyzed below.
Minor effects on the excitation energies and oscillator strengths in line with previous works~\cite{palu+09jcp,baum+12jctc1,rang+17jcp} are discussed in the Supporting Information (see Figure S6 and Table S1).
The character and spatial distribution of the electron-hole pairs are obtained from the analysis of the hole and electron densities, which for the $\lambda^{th}$ excitation are defined as~\cite{cocc+11jpcl,deco+14jpcc}
\begin{equation}
\rho_{h}^{\lambda} (\textbf{r})=\sum_{\alpha \beta} A_{\alpha \beta}^{\lambda} \rvert \phi_{\alpha}(\textbf{r})\lvert^{2}
\label{eq:h}
\end{equation}
and
\begin{equation}
\rho_{e}^{\lambda} (\textbf{r})=\sum_{\alpha \beta} A_{\alpha \beta}^{\lambda} \rvert \phi_{\beta}(\textbf{r})\lvert^{2},
\label{eq:e}
\end{equation}
respectively.
The coefficients $A_{\alpha \beta}^{\lambda}$ are the square of the normalized BSE eigenvectors that act as weighting coefficients of each transition between occupied $(\phi_{\alpha})$ and unoccupied $(\phi_{\beta})$ QP states contributing to the $\lambda^{th}$ excitation.

\section{Results and Discussion}
\begin{figure}
\centering
\includegraphics[width=0.48\textwidth]{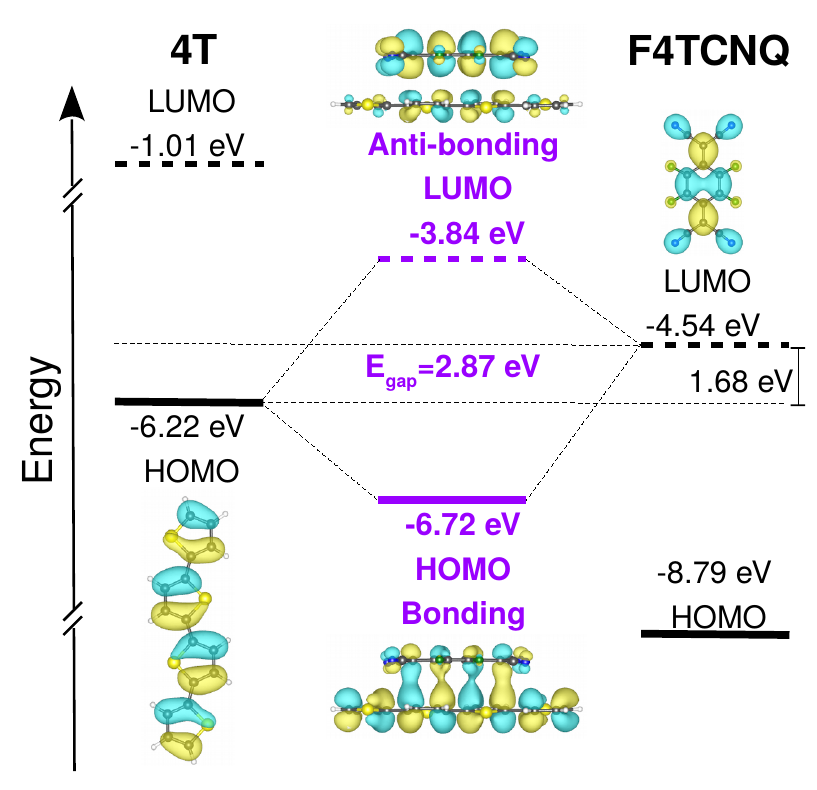}
\caption{Energy level alignment between the isolated 4T (left) and F4TCNQ (right) computed from DFT (CAM-B3LYP functional). The HOMO and the LUMO of the resulting complex  are shown in the the middle panel. For comparison, the HOMO of the donor and the LUMO of the acceptor are also depicted.}
\label{level_alignment}
\end{figure}

We start our analysis by considering the electronic structure of the CTC formed by the 4T oligomer doped by F4TCNQ, depicted in Figure~\ref{level_alignment}.
The energy values defined therein are computed with respect to the vacuum level set to zero.
By first inspecting the electronic energy levels of the individual molecules, we notice that they form a \textit{staggered} or \textit{type-II} level alignment, with the lowest-unoccupied molecular orbital (LUMO) of the acceptor lying within the gap of the donor.
In the resulting 4T-F4TCNQ complex, the highest-occupied molecular orbital (HOMO) is energetically lower than the HOMO of the donor, while the LUMO is higher than the LUMO of the acceptor (see Figure~\ref{level_alignment}, middle panel).
In agreement with the known behavior of CTCs~\cite{zhu+11cm,mend+13acie,gao+13jmcc,mend+15ncom}, the band gap of the complex is almost 1 eV larger than the energy difference between the HOMO of 4T and the LUMO of F4TCNQ.
The HOMO and LUMO of the CTC exhibit the typical features of bonding and anti-bonding states, respectively, due to the $\pi$-$\pi$ coupling between the components: The HOMO is seamlessly extended across the donor and the acceptor molecules, while the LUMO distribution is characterized by an evident nodal plane between the 4T and the F4TCNQ (see Figure~\ref{level_alignment}, middle panel).
The character of the frontier orbitals is an intrinsic property of the CTCs, independent of modeling them as dimers \textit{in vacuo}.
Bonding and anti-bonding states are formed also when the donor/acceptor complexes are represented by periodic stacks -- see Figure S5 in the Supporting Information, and Ref.~\citenum{beye+19cm}.

\begin{figure}
\centering
\includegraphics[width=0.48\textwidth]{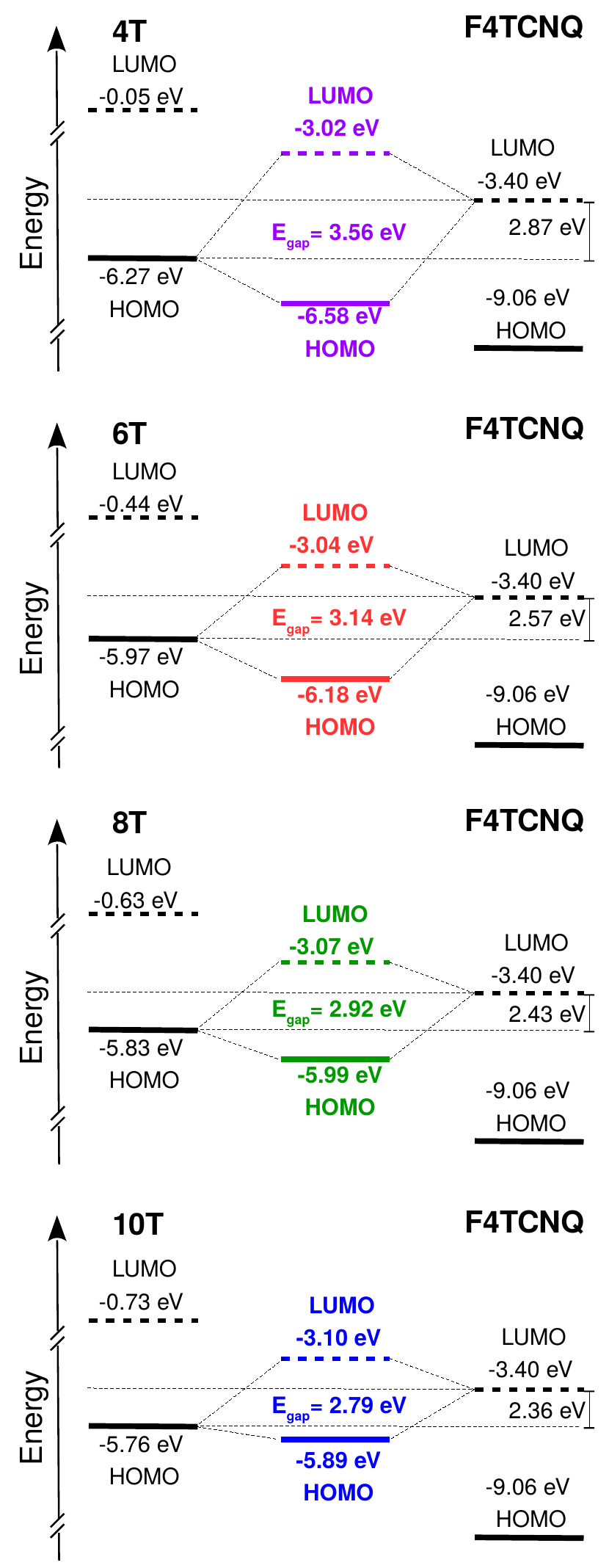}
\caption{Energy level alignment computed from $G_{0}W_{0}$ on top of DFT (CAM-B3LYP functional) for the 4T-, 6T-, 8T-, and 10T-F4TCNQ complexes and their individual components.}
\label{Elec_all}
\end{figure}

Level alignment and band gaps analyzed above are computed from DFT, employing the hybrid functional CAM-B3LYP.
An improved picture can be obtained from MBPT including the QP correction computed from $G_0W_0$.
In Figure~\ref{Elec_all} we report the results of these calculations for all the investigated systems, showing that CTCs are formed also upon increasing donor length, with the qualitative trend for the gaps and the frontier levels being unchanged compared to the case of 4T-F4TCNQ discussed above. 
As expected upon the inclusion of the electronic self-energy, the size of the band gap of the individual molecules and of the resulting CTCs is increased by about 1 eV. 
We first consider again the case of the 4T-F4TCNQ complex, shown on the top panel of Figure~\ref{Elec_all}.
In the $G_0W_0$ energy levels of both donor and acceptor separately, the LUMO is shifted up by almost 1 eV while the QP correction to the HOMO is one order of magnitude smaller and moves towards more negative energies.
As a result, the energy difference between the HOMO of the donor and the LUMO of the acceptor increases from 1.68 eV in DFT (Figure~\ref{level_alignment}) to 2.87 eV in $G_0W_0$ (Figure~\ref{Elec_all}).
The CTC exhibits a similar although less pronounced behavior, such that the QP gap of 4T-F4TCNQ is equal to 3.56 eV.
Inspecting now Figure~\ref{Elec_all} in full we notice that the QP gap decreases by about 0.8 eV going from 4T-F4TCNQ to 10T-F4TCNQ.
The band gap reduction in the CTCs is less significant than in the isolated thiophene oligomers with increasing length~\cite{fabi+05jpca,sun-auts14jctc,cocc-drax15prb}, revealing that the hybridization between donor and acceptor cannot be trivially predicted from the level alignment of the individual constituents. 
Yet, it is clear from Figure~\ref{level_alignment} that the decrease of the electronic gaps in the CTCs with varying nT size is mainly due to the up-shift of the HOMO level energy with the LUMO remaining almost constant.
As the orbital energies are computed with the vacuum level set to zero, this finding implies that the electron affinity of the CTCs is almost unaffected by the increasing donor length, while the ionization energy is critically influenced by this parameter.
This behavior is not surprising, since the examined systems differ only by the length of the donor doped by the same electron acceptor.
On the other hand, this result is not obvious \textit{a priori} considering the hybridized character of the HOMO and the LUMO in all CTCs.
From this analysis it is possible to deduce the first relevant result of this work: The increasing donor conjugation length leads to a sizable decrease in the band gaps of the CTCs (0.8 eV going from 4T-F4TCNQ to 10T-F4TCNQ) which stems almost entirely from the reduction of the ionization energy.
The physical rationale of this behavior can be deduced from the spatial distribution of the frontier orbitals (see Figure S4 in the Supporting Information). 
While the bonding and anti-bonding character of the HOMO and the LUMO is shared by all CTCs, the delocalization of the highest-occupied orbital is enhanced upon the increasing donor length.
On the contrary, even in the longest nT chains, the LUMO remains confined on the F4TCNQ and on the few thiophene rings interacting with the acceptor through $\pi$-$\pi$ coupling.
The similar trend given by $G_0W_0$ (Figure~\ref{Elec_all}) and DFT results (Figure~\ref{level_alignment} and Figure S3 in the Supporting Information) suggests that quantum confinement and electronic hybridization are mainly responsible for the variation of the gap due to the up-shift of the HOMO with almost constant LUMO energies.
Different from dimers formed by the same moiety (see \textit{e.g.}, Ref.~\citenum{hoga+13jcp}), here many-body effects, such as the electronic screening and the self-energy, improve the accuracy of orbital energies and electronic gaps, but do not qualitatively alter the picture given by (hybrid) DFT.

\begin{figure} 
\centering
\includegraphics[width=0.5\textwidth]{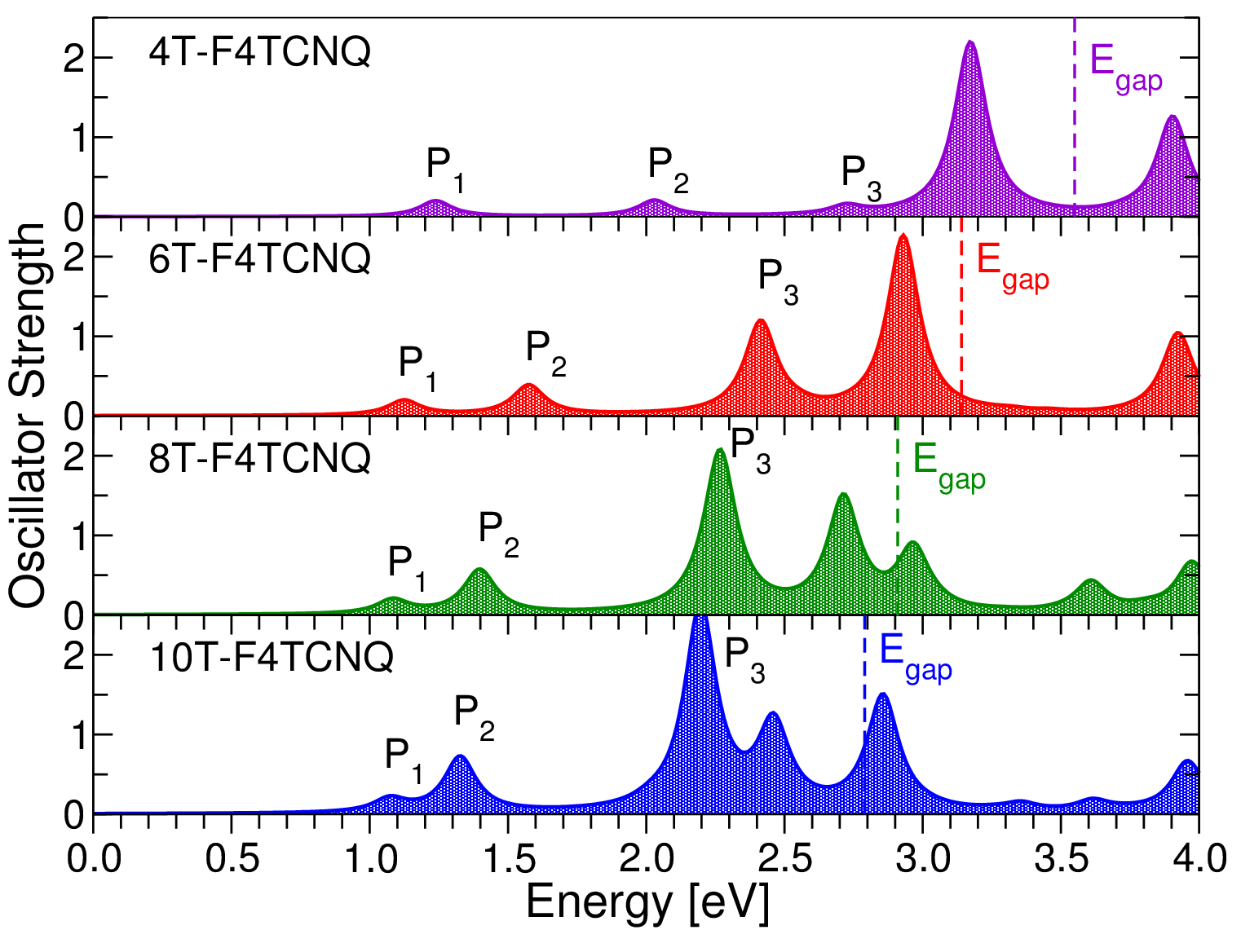}
\caption{Optical absorption spectra of the 4T-, 6T-, 8T-, and 10T-F4TCNQ complexes. The first three bright peaks are labeled as P$_{1}$, P$_{2}$, and P$_{3}$, and the energy of the QP gaps ($E_{gap}$) is marked by vertical dashed lines. }
\label{spectra}
\end{figure}

\begin{figure*} 
\centering
\includegraphics[width=1.0\textwidth]{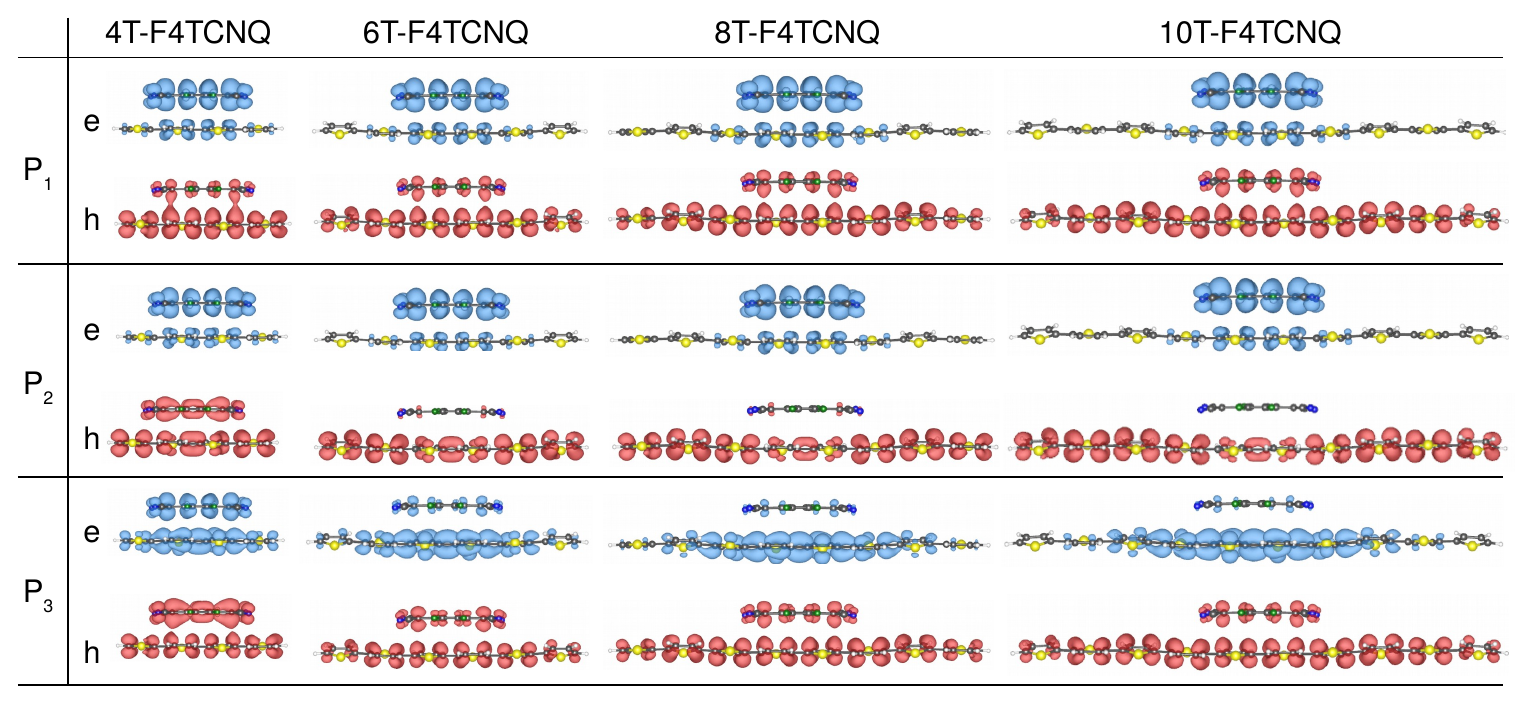}
\caption{Electron (e) and hole (h) densities of the first (P$_{1}$), second (P$_{2}$) and third (P$_{3}$) optically active excitations of each complex. Isosurfaces are plotted with a cutoff of 0.002 \AA{}$^{-3}$.}
\label{h-e}
\end{figure*}

We now turn to the optical properties of the CTCs.
In all complexes, the first excitation is dipole-allowed and gives rise to the weak peak labeled P$_1$ (see Figure~\ref{spectra}).
Regardless of the donor length, it has very similar energy ranging between 1.1 eV and 1.2 eV and comparable oscillator strength.
This behavior can be better understood by analyzing the character of this excitation in terms of orbital contributions.
In Table~\ref{CTC-composition} we report the composition of P$_1$ in the 4T-F4TCNQ and 8T-F4TCNQ complexes (details about the other CTCs are given in the Supporting Information, Table S2) and in Figure~\ref{h-e} we display the spatial distribution of the corresponding electron and hole densities (Eqs.~\ref{eq:h} and~\ref{eq:e}).
In all complexes the first excitation stems from the HOMO-LUMO transition, such that the hole and the electron components of P$_1$ exhibit the same character and spatial extension of the frontier orbitals.
This has two main consequences: In all systems the lowest-energy excitation has a Frenkel-like character localized within the CTC. 
Moreover, the comparable overlap between the HOMO and the LUMO (and equivalently of the hole and the electron density) in all complexes explains why P$_1$ has almost constant oscillator strength going from 4T-F4TCNQ to 10T-F4TCNQ.
We also recall that P$_1$ retains approximately the same energy in all CTCs irrespective of the donor length, in contrast with the decrease of the electronic gaps in the presence of longer nT chains discussed above. 
To clarify this point, we take advantage of the adopted formalism where optical excitations are computed including the electron-hole correlation.
In this framework, the difference between the quasi-particle HOMO-LUMO gap and the excitation energy of P$_1$ can be interpreted as the exciton binding energy ($E_b$).
By comparing the excitation energies in Table~\ref{CTC-composition} (and Table S2 in the Supporting Information) with the QP gaps in Figure~\ref{Elec_all}, it is evident that $E_b$ decreases as the length of the donor oligomer increases.
The binding energy reduction of P$_1$ upon increasing donor length is therefore comparable with the decrease of the band gap (see Figure~\ref{Elec_all}), such that these two quantities end up compensating each other.
In the presence of longer nT chains the Coulomb screening is enhanced and the quantum confinement of the electron-hole pairs is concomitantly reduced.
The Coulomb screening is solely embedded in the direct term of the BSE while quantum confinement affects also the bare exchange interaction between the electron and the hole~\cite{hoga+13jcp,cocc-drax15prb}. 
Due to these intertwined effects and to the reduction of the QP gaps discussed above, P$_1$ appears at the same energy in all the examined CTCs, regardless of the donor conjugation length.

\begin{table*} 
\centering
\caption{Energies, oscillator strength (OS), and composition in terms of transitions between molecular orbitals of the first three bright excitations in the 4T-F4TCNQ and 8T-F4TCNQ complexes. HOMO and LUMO are abbreviated by H and L, respectively.}
\label{CTC-composition}%
\begin{tabular}{lcccl}
\hline
& Excitation & Energy [eV]& OS &  Composition \\ \hline
4T-F4TCNQ   & P$_1$  & 1.23 & 0.19& H $\rightarrow$ L \\ 
& P$_2$ & 2.02 & 0.19  & H-1 $\rightarrow$ L \\ 
& P$_3$  & 2.72 & 0.10 & H $\rightarrow$ L+1  \\         
& & & & H-2 $\rightarrow$ L \\ \hline 
8T-F4TCNQ & P$_1$ & 1.08 & 0.17 & H $\rightarrow$ L  \\ 
& P$_2$ & 1.39 & 0.55 & H-1 $\rightarrow$ L \\ 
& P$_3$ & 2.26 & 2.00 & H $\rightarrow$ L+1 \\ \hline	\end{tabular}
\end{table*}

Above the absorption onset a number of peaks appears in all the spectra shown in Figure~\ref{spectra}.
In the following, we focus on the second and third bright excitations, P$_2$ and P$_3$, and analyze their behavior. 
In the spectrum of 4T-F4TCNQ both P$_2$ and P$_3$ have comparable oscillator strength with respect to each other and to P$_1$. 
The electron and hole densities of these three excitations are indeed very similar, corresponding in all cases to Frenkel excitons localized within the CTC (see Figure~\ref{h-e}). 
Their different composition in terms of single QP transitions (see Table~\ref{CTC-composition}) reveals that in this complex, where the sizes of the donor and the acceptor are comparable, also the HOMO-1 and the LUMO+1 are hybridized (further details in the Supporting Information, see Figure S4).
This picture drawn by our first-principles results is consistent with the experimental data available for the 4T-F4TCNQ complex.
In Ref.~\citenum{mend+15ncom} the optical density of this CTC is reported: The sharp maximum at the absorption onset is followed by a dip and by a broader and more intense peak around 2.4 eV. 
Two findings are particularly significant: First, the energy separation between the first and second maxima in the measured spectra is consistent with the computed energy difference between P$_1$ and P$_3$ shown in Figure~\ref{spectra} and Table~\ref{CTC-composition}. 
Second, the relative intensity between the two maxima recorded in the experiment follows the same trend resulting in the calculated spectrum. 
Absolute values of excitation energies and binding energies are not quantitatively comparable between our theoretical results and the experiment, as the CTCs are modeled here by dimers \textit{in vacuo}. 
Thus, only relative trends can be correlated.
We also note that in organic films the exciton binding energy may be affected by the crystalline structure of the doped material, which in turns may influence the hole delocalization -- see, \textit{e.g.}, Ref.~\citenum{li+17prm}.

The scenario changes substantially when the CTCs are formed by oligothiophenes with six rings or more: P$_2$ and P$_3$ have significantly higher oscillator strength than P$_1$ (see Figure~\ref{spectra}) and their character varies accordingly (see Figure~\ref{h-e}). 
The orbital composition of these two excitations is analogous to the case of 4T-F4TCNQ (see Table~\ref{CTC-composition}), demonstrating that the influence of the donor length manifests itself primarily in the character of the orbitals.
In these CTCs with enhanced size mismatch between the donor and the acceptor the orbitals below the HOMO and above the LUMO remain localized only on the nT oligomer (see Supporting Information, Figure S4): Being considerably longer than F4TCNQ, 6T, 8T, and 10T have higher density of states around the gap. 
As a consequence, P$_2$, which stems from the transition between the HOMO-1 and the LUMO in all CTCs, becomes a charge-transfer exciton with the hole localized on the donor and the electron almost completely spread on the acceptor. 
The enhanced space separation between electron and hole densities is responsible for the relative low oscillator strength of P$_2$ compared to higher-energy peaks.
On the other hand, both the electron and the hole in P$_3$ tend to be increasingly localized on the nT chain as the size of the latter grows, giving rise to a Frenkel exciton within the donor. 
The significant increase of the oscillator strength of P$_3$ upon increasing length of the thiophene oligomers is therefore explained by the larger overlap between the electron and the hole components of this excitation. 
Finally, by inspecting all in all the spectra shown in Figure~\ref{spectra}, we notice that different from P$_1$, but in line with the behavior of the electronic gaps, both P$_2$ and P$_3$ red-shift upon increasing donor length and so do also the peaks at higher energies.
The concomitant rise of their oscillator strength results in increased spectral density in the visible region.
The relevance of this result is apparent: Even in case only partial charge transfer occurs in doped organic semiconductors, photo-excited charge carriers can be effectively generated upon efficient absorption of visible light and enhanced electron-hole separation.

Before concluding, it is instructive to discuss the trends obtained by our results in light of the available experimental findings on nT-F4TCNQ charge-transfer complexes.
The optical spectrum measured for 4T-F4TCNQ with 50$\%$ doping concentration and reported in Ref.~\citenum{mend+15ncom} is dominated by a bright peak at the onset which is interpreted as a signature of CTC formation.
The character of $P_1$ discussed above is fully compatible with this analysis.
Moreover, the experimental spectrum is characterized by a sharp increase of the oscillator strength about 2 eV above the optical gap, which is also in line with the results shown in Figure~\ref{spectra}.
Due to the lack of measurements on CTCs formed by longer oligomers than 4T, it is not possible to extend this comparison to the other systems investigated in this work.
However, considering the tendency of P3HT to bend, forming linear units consisting of at least half a dozen of thiophene rings, it is reasonable to discuss our results in light of measurements performed on P3HT-F4TCNQ blends where partial charge transfer is locally observed~\cite{jaco+18mh,neel+18jpcl}.
The measured spectra reveal a broad absorption band with a maximum between 2.0 eV and 2.5 eV, which is assigned to transitions between hybridized states and thus interpreted as a signature of CTC formation~\cite{jaco+18mh,neel+18jpcl}. 
This feature can be related to the peak P$_3$ in the spectra of 8T-F4TCNQ and 10T-F4TCNQ shown in Figure~\ref{spectra}, taking into account the expected blue-shift of a few hundred meV compared to experiments induced by modelling the CTCs as dimers \textit{in vacuo}.
New experiments on nT-F4TCNQ CTCs with longer donor chains than 4T are certainly needed to corroborate this analysis.
A quantitative comparison with future measurements will be enabled by the inclusion in the calculated spectra of an effective screening term accounting for the polarizable environment surrounding the doped species in the crystalline phase.

\section{Summary and Conclusions}
In summary, we have investigated from first-principles many-body theory how the donor conjugation length impacts the electronic structure and the optical excitations of CTCs formed by thiophene oligomers doped by the strong electron acceptor F4TCNQ.
We have found that all systems are characterized by hybridized frontier orbitals regardless of the size of the thiophene oligomers. 
The electronic gaps decrease upon increasing donor length solely due to the reduction of the ionization energy, with the electron affinity remaining almost constant in all CTCs. 
The optical gaps follow a different trend, being at approximately the same energy irrespective of the donor length.
Together with the comparable oscillator strength of the first absorption peak, this finding implies that whenever a nT-F4TCNQ charge-transfer complex is formed, it absorbs the same about of light at approximately the same energy independently of the donor length. 
The size of the nT oligomer influences the strength and character of the higher-energy excitations. 
While in the 4T-F4TCNQ complex the first three bright excitations retain the same oscillator strength, in the presence of thiophene oligomers with six rings or more, higher-energy peaks gain significant spectral weight relative to the absorption onset and experience a sizable red-shift.
As a result, the optical density in the visible region is significantly enhanced upon increasing donor length.
The nature of the excitations changes accordingly. 
In 4T-F4TCNQ the first three excitations have the same Frenkel-like character within the CTC.
Conversely, in the complexes formed by longer nT chains the second bright peak corresponds to a charge-transfer-like excitation and the third one to a Frenkel exciton with increasing localization on the donor as the size of the latter grows.

Our work and the rationale elaborated therein offer important insight into the electronic structure and optical excitations of charge-transfer complexes.
While doping in organic semiconductors remains a debated topic, our findings unravel fundamental structure-property relations in the presence of partial charge transfer, which is one of the leading doping mechanisms in molecular materials.
As such, our results represent an important step forward towards a deeper understanding of the opto-electronic properties of doped organic semiconductors.

\begin{acknowledgement}
Inspiring discussions with Arrigo Calzolari, Andreas Optiz, and Dieter Neher are gratefully acknowledged. 
This work is funded by the Deutsche Forschungsgemeinschaft (DFG, German Research Foundation) - Projektnummer 182087777 - SFB 951 and HE 5866/2-1.
\end{acknowledgement}


\begin{suppinfo}
Additional details about the structural, electronic, and optical properties of the CTCs are reported in the Supporting Information.
\end{suppinfo}
\newpage

\section*{Graphical TOC Entry}
\begin{figure}%
\centering
\includegraphics[height=4.45cm]{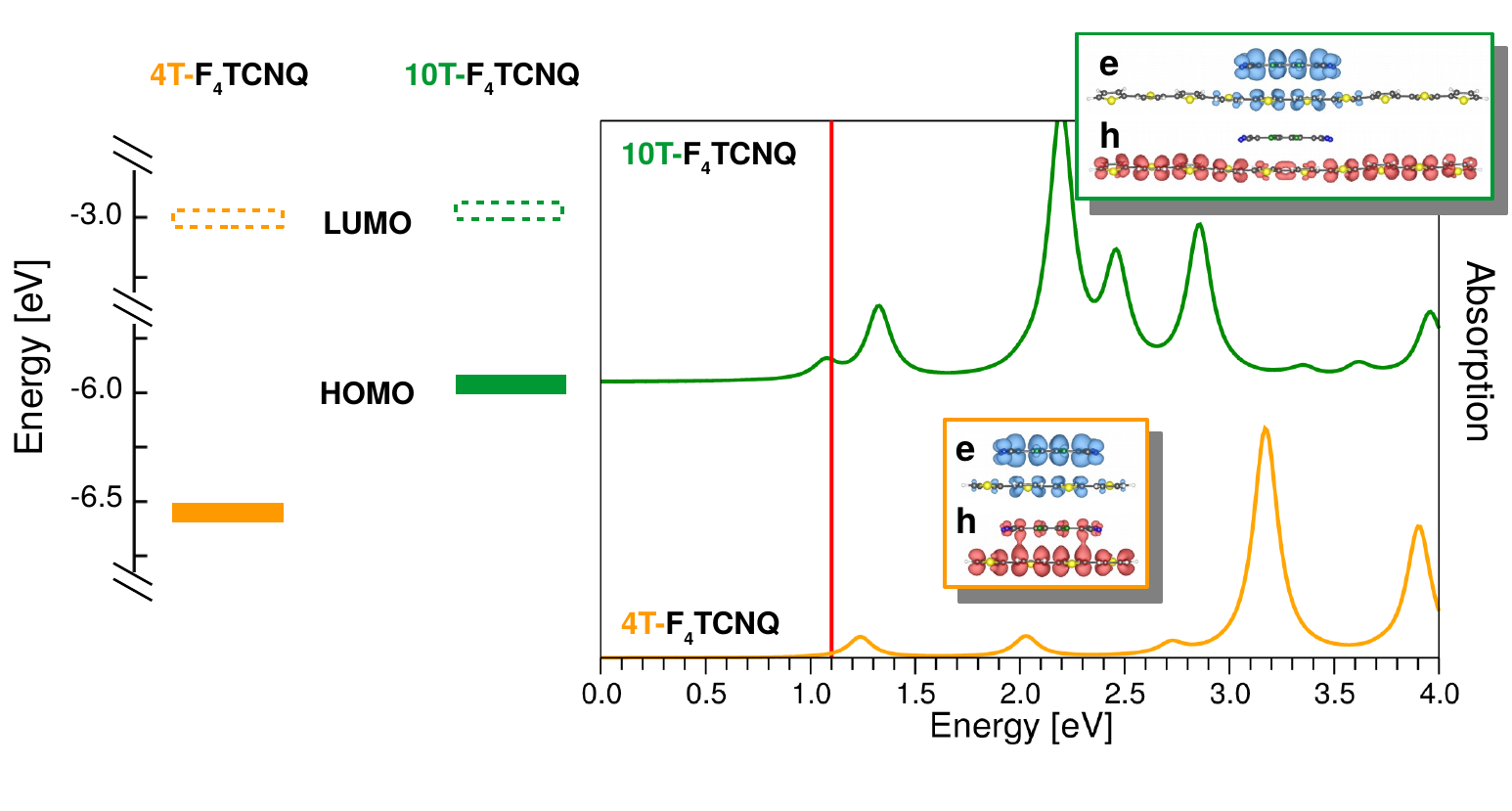}
\end{figure}
\newpage
\providecommand{\latin}[1]{#1}
\providecommand*\mcitethebibliography{\thebibliography}
\csname @ifundefined\endcsname{endmcitethebibliography}
  {\let\endmcitethebibliography\endthebibliography}{}


\end{document}